\documentclass{article}

\usepackage[numbers]{natbib}

\usepackage[preprint]{neurips_2024}

\usepackage[utf8]{inputenc} 
\usepackage[T1]{fontenc}    
\usepackage{hyperref}       
\usepackage{url}            
\usepackage{booktabs}       
\usepackage{amsfonts}       
\usepackage{nicefrac}       
\usepackage{microtype}      
\usepackage{xcolor}         

\usepackage[utf8]{inputenc}
\usepackage{geometry}

\usepackage{graphicx}
\usepackage{amsmath}
\usepackage{amssymb}
\usepackage{booktabs}
\usepackage{listings}
\usepackage{xcolor}
\usepackage{hyperref,authblk}
\usepackage{float}
\usepackage{enumitem}

\usepackage{caption}
\usepackage{tcolorbox}
\usepackage{times}
\usepackage[T1]{fontenc}

\title{DIAP: A Decentralized Agent Identity Protocol with Zero-Knowledge Proofs and a Hybrid P2P Stack}

\author[1]{Yuanjie Liu}
\author[1,2]{Wenpeng Xing}
\author[4]{Ye Zhou}
\author[4]{Gaowei Chang}
\author[1,3]{Changting Lin}
\author[2,3]{Meng Han}

\affil[1]{\textit{Binjiang Institute of Zhejiang University}}
\affil[2]{\textit{Zhejiang University}}
\affil[3]{\textit{GenTel.io}}
\affil[4]{\textit{ANP Open Source Community}}

\hypersetup{
    colorlinks=true,
    linkcolor=blue!70!black,
    citecolor=green!60!black,
    urlcolor=blue!80!black,
    pdftitle={DIAP: A Decentralized Agent Identity Protocol},
    pdfauthor={liuyuanjie},
    pdfsubject={Computer Science, Cryptography, Networking},
    pdfkeywords={DIAP, Agent, Decentralized Identity, ZKP, Noir, IPFS, Libp2p, Iroh, Rust}
}

\definecolor{codegray}{rgb}{0.5,0.5,0.5}
\definecolor{codepurple}{rgb}{0.58,0,0.82}
\definecolor{codeblue}{rgb}{0,0,0.6}
\definecolor{codegreen}{rgb}{0.1,0.5,0.1} 
\definecolor{backcolor}{rgb}{0.98,0.98,0.98}

\lstdefinestyle{ruststyle}{
    language=Rust,
    backgroundcolor=\color{backcolor},
    commentstyle=\color{codegreen},
    keywordstyle=\color{codeblue},
    numberstyle=\tiny\color{codegray},
    stringstyle=\color{codepurple},
    basicstyle=\ttfamily\footnotesize,
    breakatwhitespace=false,
    breaklines=true,
    captionpos=b,
    keepspaces=true,
    numbers=left,
    numbersep=5pt,
    showspaces=false,
    showstringspaces=false,
    showtabs=false,
    tabsize=2
}

\lstdefinestyle{noirstyle}{
    language=[Objective-C]C, 
    backgroundcolor=\color{backcolor},
    commentstyle=\color{codegreen},
    keywordstyle=\color{codeblue},
    morekeywords={fn, main, assert, pub, Field}, 
    numberstyle=\tiny\color{codegray},
    stringstyle=\color{codepurple},
    basicstyle=\ttfamily\footnotesize,
    breakatwhitespace=false,
    breaklines=true,
    captionpos=b,
    keepspaces=true,
    numbers=left,
    numbersep=5pt,
    showspaces=false,
    showstringspaces=false,
    showtabs=false,
    tabsize=2
}

\lstdefinestyle{noirstyle}{
    language=Noir,
    basicstyle=\ttfamily\small,
    keywordstyle=\color{blue},
    commentstyle=\color{gray},
    stringstyle=\color{orange},
    numbers=left,
    numberstyle=\tiny,
    stepnumber=1,
    numbersep=5pt,
    tabsize=4,
    showstringspaces=false,
    breaklines=true,
}

\lstdefinelanguage{Noir}{
    morekeywords={fn, let, pub, struct, impl, for, return, if, else, match, use, assert},
    sensitive=true,
    morecomment=[l]{//},
    morestring=[b]",
}

\begin{document}

\maketitle

\begin{abstract}
The absence of a fully decentralized, verifiable, and privacy-preserving communication protocol for autonomous agents remains a fundamental challenge in decentralized computing. Existing frameworks either rely on centralized intermediaries—reintroducing trust bottlenecks—or fail to integrate decentralized identity-resolution mechanisms , forfeiting advantages in persistence and cross-network resolvability.
To address this limitation, we propose the Decentralized Interstellar Agent Protocol (DIAP) — a novel framework for agent identity and communication that achieves persistent, verifiable, and trustless interoperability in fully decentralized environments. DIAP binds an agent’s identity to an immutable IPFS/IPNS Content Identifier (CID) and employs Zero-Knowledge Proofs (ZKP) to dynamically and statelessly prove ownership, eliminating the need for record updates.
We further present a Rust SDK integrating Noir (ZKP), DID:Key, IPFS, and a hybrid P2P stack combining Libp2p GossipSub for discovery and Iroh for high-performance, QUIC-based data exchange. Through a “zero-dependency” ZKP deployment model with a precompiled Noir circuit, DIAP enables instant, verifiable, and privacy-preserving identity proof—establishing a practical foundation for next-generation autonomous agent ecosystems.
Furthermore, the SDK introduces a significant engineering innovation: a "zero-dependency" ZKP deployment model. Through a universal ZKP manager and a sophisticated compile-time build script, the SDK embeds a pre-compiled Noir circuit, removing the requirement for end-users to install any external ZKP compilers or toolchains. This work provides a practical, high-performance, and privacy-preserving foundation for the next generation of autonomous agent-to-agent (A2A) economies. The code is available at \href{https://github.com/logos-42/DIAP_Rust_SDK}{https://github.com/logos-42/DIAP\_Rust\_SDK}.
\end{abstract}

\section{Introduction}
As autonomous agents increasingly mediate digital and physical decision-making—across Large Language Models (LLMs), Decentralized Physical Infrastructure Networks (DePIN), and Internet of Things (IoT) ecosystems—the absence of a trust foundation among agents has become a critical bottleneck. Without a verifiable, persistent, and decentralized identity layer, inter-agent communication reverts to centralized trust anchors, undermining auditability, autonomy, and composabilit.
In blockchain-driven ecosystems, where agents must independently verify ownership, execute transactions, and coordinate actions without intermediaries, this lack of a native trust substrate becomes even more restrictive.

Existing attempts—such as centralized key registries or permissioned agent networks—are only temporary transitional solutions and cannot fully release agent’s potential.

However,current solutions face a critical dilemma. Centralized identity providers (IdPs) create single points of failure and control, antithetical to the goals of a decentralized web. Despite the prominence of existing agent communication protocols such as A2A and ACP, current solutions face a series of structural inefficiencies and conceptual contradictions. Centralized identity providers (IdPs) still introduce single points of failure and control, fundamentally conflicting with the ethos of decentralization. 

Furthermore, while ACP has been accepted by blockchain ecosystems, its integration remains cumbersome, plagued by complex intermediations that limit efficiency and blockchain compatibility. Similarly, A2A protocols, though theoretically elegant, often operate within centralized paradigms that hinder large-scale, verifiable collaboration among agents. In addition, existing frameworks struggle to balance collaborative extensibility with privacy protection, restricting both the scope and granularity of agent cooperation. Finally, most current protocols lack lightweight accessibility and developer friendliness.

Consequently, these contradictions underscore not merely a technical gap, but a foundational absence of a unifying paradigm capable of reconciling decentralization, scalability, verifiable trust, and efficient interoperability among autonomous agents. Existing frameworks either sacrifice decentralization for performance or compromise privacy for verifiability—an inherent paradox that has constrained the evolution of autonomous agent ecosystems.

The Decentralized Interstellar Agent Protocol (DIAP) therefore introduces a new foundational layer for the post-centralized web: a cryptographically verifiable, privacy-preserving, and blockchain-native communication substrate that redefines how intelligent agents establish identity, prove ownership, and cooperate autonomously across trustless environments.

In this paper, we present the Decentralized Interstellar Agent Protocol (DIAP), a novel approach that fundamentally rethinks agent identity. DIAP's core premise is: Identity should be immutable, while proof of ownership should be dynamic and stateless.

Instead of relying on mutable pointers, DIAP binds an agent's identity to the immutable, cryptographic hash of its identity document: its IPFS CID. To solve the problem of "identity revocation" or "key rotation" (which mutable pointers solve clumsily), DIAP employs a more powerful cryptographic primitive: Zero-Knowledge Proofs (ZKP). An agent proves its continued ownership of an identity by generating a ZKP that attests---without revealing its private key---that it controls the cryptographic keys associated with the public `DIDDocument` stored at that CID.

This report details the design and implementation of the DIAP Rust SDK, a full-featured, production-ready library for building DIAP-compatible agents. Our contributions are fourfold:

\begin{enumerate}
    \item \textbf{Protocol Design (ZKP-on-CID):} A novel identity protocol that uses an immutable IPFS CID as a permanent agent identifier and Noir-based ZKP for stateless, instantaneous proof of ownership.
    \item \textbf{Hybrid P2P Network Stack:} A dual-stack communication model combining Libp2p GossipSub for authenticated service discovery and Iroh for high-performance, direct agent interaction.
    \item \textbf{Privacy-Preserving Architecture:} An end-to-end privacy model featuring cryptographic signing of messages and encryption of P2P network identifiers (`EncryptedPeerID`) to prevent network-level reconnaissance.
    \item \textbf{Zero-Dependency ZKP SDK:} A novel engineering solution (`UniversalNoirManager`) that uses Rust's build-script system (`build.rs`) to pre-compile and embed the ZKP circuit, delivering a zero-dependency, cross-platform SDK for developers.
\end{enumerate}

\begin{figure}[t]
    \centering
\includegraphics[width=1\linewidth]{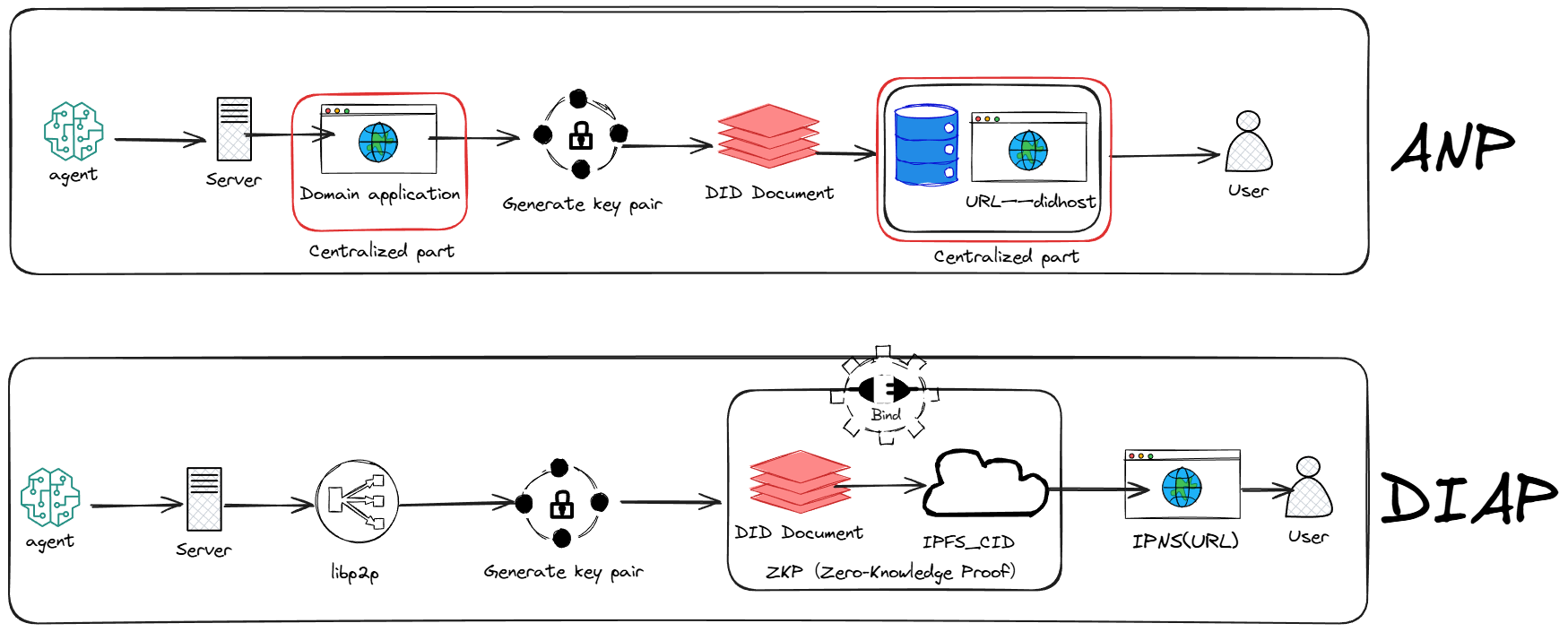}
    \caption{Comparison between ANP and DIAP.}
    \label{fig:placeholder}
\end{figure}
The InterPlanetary Naming System (IPNS) extends IPFS by introducing mutable naming capabilities on top of immutable content addressing. While IPFS provides cryptographically verifiable permanence through Content Identifiers (CIDs), IPNS allows these CIDs to be dynamically updated via cryptographic keys—enabling references to evolve without changing the underlying address. In principle, this makes IPNS a candidate for decentralized identity resolution and agent discoverability. However, due to its design reliance on cryptographic key republishing and DHT-based propagation, IPNS has often been dismissed as unsuitable for high-frequency, low-latency applications.

Nevertheless, it is crucial to recognize that the very characteristics that constrain IPNS in high-frequency, low-latency contexts—its reliance on mutable pointers and periodic republishing—also underpin its unique strength as a persistent identity layer. By anchoring mutable states to immutable addressing, IPNS achieves a rare equilibrium between flexibility and stability, providing a consistent interface for decentralized identity resolution. In this sense, what appears as inefficiency under real-time communication constraints becomes paradoxically a virtue for long-term identity persistence and verifiable agent discoverability.

\section{System Architecture}
The DIAP SDK is architected as a modular, layered stack. Each layer provides a distinct set of functionalities, allowing agents to establish identity, prove ownership, discover peers, and communicate securely.

\begin{figure}[t]
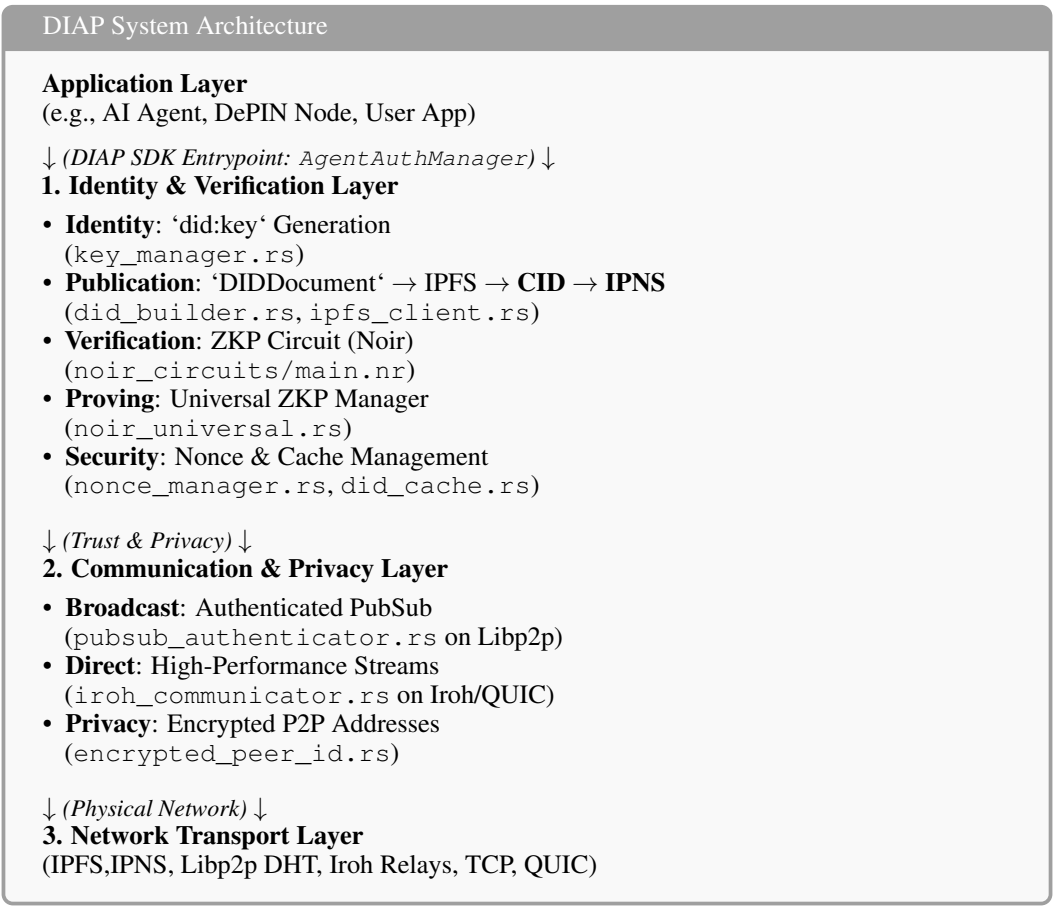

    \centering
    \begin{tcolorbox}[width=\columnwidth, colback=gray!5, colframe=gray!75, title=DIAP System Architecture]
    \textbf{Application Layer} \\
    (e.g., AI Agent, DePIN Node, User App)
    \vspace{2mm}
    
    $\downarrow$ \textit{\footnotesize (DIAP SDK Entrypoint: \texttt{AgentAuthManager})} $\downarrow$
    
    \textbf{1. Identity \& Verification Layer}
    \begin{itemize}[leftmargin=*,noitemsep]
        \item \textbf{Identity}: `did:key` Generation \\ (\texttt{key\_manager.rs})
        \item \textbf{Publication}: `DIDDocument` $\rightarrow$ IPFS $\rightarrow$ \textbf{CID} $\rightarrow$ \textbf{IPNS}\\ (\texttt{did\_builder.rs}, \texttt{ipfs\_client.rs})
        \item \textbf{Verification}: ZKP Circuit (Noir) \\ (\texttt{noir\_circuits/main.nr})
        \item \textbf{Proving}: Universal ZKP Manager \\ (\texttt{noir\_universal.rs})
        \item \textbf{Security}: Nonce \& Cache Management \\ (\texttt{nonce\_manager.rs}, \texttt{did\_cache.rs})
    \end{itemize}
    \vspace{2mm}
    
    $\downarrow$ \textit{\footnotesize (Trust \& Privacy)} $\downarrow$
    
    \textbf{2. Communication \& Privacy Layer}
    \begin{itemize}[leftmargin=*,noitemsep]
        \item \textbf{Broadcast}: Authenticated PubSub \\ (\texttt{pubsub\_authenticator.rs} on Libp2p)
        \item \textbf{Direct}: High-Performance Streams \\ (\texttt{iroh\_communicator.rs} on Iroh/QUIC)
        \item \textbf{Privacy}: Encrypted P2P Addresses \\ (\texttt{encrypted\_peer\_id.rs})
    \end{itemize}
    \vspace{2mm}
    
    $\downarrow$ \textit{\footnotesize (Physical Network)} $\downarrow$
    
    \textbf{3. Network Transport Layer} \\
    (IPFS,IPNS, Libp2p DHT, Iroh Relays, TCP, QUIC)
    \end{tcolorbox}
    \caption{The layered architecture of the DIAP SDK.}
    \label{fig:architecture}
\end{figure}

The flow is as follows: An agent first uses the Identity Layer to generate keys and publish a \verb|DIDDocument| to IPFS, receiving an immutable CID. When interacting, it uses the Verification Layer to generate a ZKP. This proof, along with its message, is sent over the Communication Layer, which multiplexes between a Libp2p broadcast network and an Iroh direct connection network. Furthermore, the CID can be bound to an IPNS record, enabling persistent and mutable address registration. Subsequent modifications to the \verb|DIDDocument| can be automatically verified via ZKP against the previous CID, with the updated CID seamlessly propagated through IPNS, thereby preserving both integrity and continuity within the decentralized identity namespace.

\section{Identity \& Verification Layer}
Unlike traditional identity frameworks that tightly couple stateful identity with verification logic, DIAP’s core innovation lies in decoupling immutable identity anchors from transient proofs, thereby enabling secure, low-latency interactions without sacrificing verifiability. 

\subsection{Identity Registration: ZKP-on-CID}
The identity registration process is defined in \path{src/did_builder.rs} and \path{src/key_manager.rs}. It is an offline, one-time operation.

\begin{enumerate}
    \item \textbf{Key Generation:} The agent generates a standard \texttt{ed25519} keypair (\texttt{KeyPair}) using \texttt{src/key\_manager.rs}.
    \item \textbf{DID Derivation:} From the public key, a W3C-compliant \texttt{did:key} identifier is derived (e.g., \texttt{did:key:z6Mk...}).
    \item \textbf{Document Creation:} A \texttt{DIDDocument} (JSON) is created. This document contains the agent's public key, its \texttt{did:key} as the controller, and, critically, its P2P network information (see Sec 3.2).
\item \textbf{Publication:} This \texttt{DIDDocument} is uploaded to IPFS via \texttt{src/ipfs\_client.rs}.
\item \textbf{Identity Finalization:} The protocol returns the immutable IPFS \textbf{CID} (e.g., \texttt{Qm...}) of the document.
\item \textbf{Persistent Identity Binding:} The obtained \textbf{CID} is subsequently linked to an \textbf{IPNS} record, creating a stable and human-readable identity gateway for external discovery. This binding allows agents to maintain a consistent external identity while internally rotating cryptographic keys or updating state data, achieving persistent, verifiable, and trust-minimized identity exposure.

\end{enumerate}

This \textbf{IPNS} is now the agent's permanent, globally unique, and verifiable identity. 
The agent's "identity" is the verifiable claim that it possesses knowledge of the binding between the \textbf{DID document} and its corresponding \textbf{CID}. 
If desired, the protocol also permits direct verification using the immutable \textbf{CID} alone, allowing agents to operate without IPNS binding while maintaining full cryptographic integrity and trustless verifiability.

\subsection{Privacy-Preserving P2P Endpoints}
A naive implementation would expose the agent's P2P network address (its \texttt{PeerId}) in plain text within the public \texttt{DIDDocument}. 
While superficially transparent, this approach is in fact self-defeating: by prioritizing openness, it annihilates security. 
Any adversary could trivially crawl IPFS, reconstruct the global agent topology, and orchestrate precise DDoS or reconnaissance attacks—turning the very transparency that was meant to ensure trust into the source of systemic vulnerability. 
\textbf{By sharp contrast}, DIAP introduces a novel \textit{self-encryption} mechanism that inverts this paradigm. 
Rather than exposing connectivity as a public attribute, each agent cryptographically encapsulates its own network coordinates, making them verifiable yet undiscoverable. 
In doing so, DIAP transforms the fundamental tension between transparency and confidentiality into a new cryptographic equilibrium—where agents remain globally identifiable without ever being globally exposed.
DIAP solves this using a novel self-encryption scheme, implemented in 
\path{src/encrypted_peer_id.rs}.

\begin{enumerate}
    \item During identity creation, the agent's \texttt{PeerId} is encrypted.
    \item The encryption key is not a new, separate key. Instead, an AES-256 key is derived from the agent's own \texttt{ed25519} private key and a static salt (e.g., \texttt{b"DIAP\_AES\_KEY\_V3"}).
    \item The resulting ciphertext, a random nonce, and a signature (over the ciphertext and nonce) are stored in the \texttt{DIDDocument} as the \texttt{serviceEndpoint}.
\end{enumerate}

This design has a powerful asymmetrical property:
\begin{itemize}
    \item \textbf{Public Verifiers} can download the \texttt{DIDDocument} and validate the \textit{signature} of the encrypted data using the agent's public key. This proves the data was not tampered with.
    \item \textbf{Only the Agent Itself} (the holder of the private key) can derive the AES key and \textit{decrypt} its own \texttt{PeerId} to establish network connections.
\end{itemize}

\subsection{The Noir ZKP Circuit}
The protocol's core verification logic is implemented as a ZKP circuit in the Noir language 
(\path{noir_circuits/src/main.nr}), shown in Figure~\ref{fig:noir_circuit}.

This circuit allows an agent (Prover) to convince a Verifier of a composite statement:

\begin{tcolorbox}[colback=gray!5, colframe=gray!75, title=DIAP ZKP Attestation]
\textit{``I possess a secret key \texttt{sk} and a secret document \texttt{doc} such that:''}

\begin{enumerate}[label=\arabic*.]
    \item The hash of \texttt{DID\_doc} matches the public \texttt{CID\_hash}.
    \item \texttt{sk} correctly derives the public key \texttt{PK\_hash} found in \texttt{doc}.
    \item I know a secret \texttt{nonce} that hashes to the public \texttt{Nonce\_hash} provided by the Verifier.
\end{enumerate}
\end{tcolorbox}

\begin{figure}[t]
\begin{lstlisting}[style=noirstyle, 
    caption={The core DIAP ZKP circuit logic from $noir\_circuits/src/main.nr$.}, 
    label=fig:noir_circuit
    ]
fn main(
    // Public inputs (known to verifier)
    expected_did_hash: [Field; 2],
    public_key_hash: Field,
    nonce_hash: Field,
    
    // Private inputs (secret witness)
    secret_key: [Field; 2],
    did_document_hash: [Field; 2],
    nonce: [Field; 2]
) -> pub Field {
    
    // Constraint 1: Verify DID document hash
    assert(did_document_hash[0] == expected_did_hash[0]);
    assert(did_document_hash[1] == expected_did_hash[1]);
    
    // Constraint 2: Verify key derivation
    let derived_key_hash = secret_key[0] * secret_key[1] + 
                           secret_key[0] + 
                           secret_key[1];
    assert(derived_key_hash == public_key_hash);
    
    // Constraint 3: Verify nonce binding
    let computed_nonce_hash = nonce[0] * nonce[1] + 
                              nonce[0] + nonce[1];
    assert(computed_nonce_hash == nonce_hash);
    
    // Constraint 4: Integrity binding
    let binding_proof = (secret_key[0] + secret_key[1]) * 
                        (did_document_hash[0] + did_document_hash[1]) +
                        nonce[0] + nonce[1];
    
    binding_proof
}
\end{lstlisting}
\end{figure}

The \texttt{nonce} (Number used once) is provided by the Verifier as a challenge and managed by \texttt{src/nonce\_manager.rs} to prevent replay attacks. This ZKP mechanism establishes a fast, stateless, and privacy-preserving verification layer that strengthens decentralized identity authentication without relying on any persistent naming or stateful identity registry. 
\section{Hybrid P2P Communication Stack}
Recognizing that heterogeneous agent interactions impose diverse latency, reliability, and topology requirements, DIAP introduces a hybrid communication stack that strategically integrates \textbf{two parallel and interoperable P2P frameworks}. This design enables agents to dynamically optimize between high-throughput gossip dissemination and low-latency, verifiable point-to-point coordination, thereby achieving a balance between scalability and trust integrity rarely attainable in conventional decentralized systems. 

\subsection{Libp2p GossipSub for Discovery}
For service discovery and N-to-N (many-to-many) broadcast, DIAP employs \texttt{libp2p-gossipsub}. A naive gossipsub implementation is insecure, as any participant can spoof messages.

DIAP solves this by wrapping all messages in a verifiable envelope, as implemented in \texttt{src/pubsub\_authenticator.rs}. The \texttt{AuthenticatedMessage} struct is a critical data structure that enforces security on the gossip layer. It requires every message to include:
\begin{itemize}
    \item \texttt{from\_did}: The sender's \texttt{did:key}.
    \item \texttt{did\_cid}: The sender's IPFS CID identity.
    \item \texttt{nonce}: A unique nonce to prevent replay.
    \item \texttt{zkp\_proof}: A valid ZKP proving ownership of the \texttt{did\_cid}.
    \item \texttt{signature}: A signature over the message content, nonce, and topic, created by the private key.
\end{itemize}
When a node receives a message (see \texttt{examples/pubsub\_verification\_loop\_demo.rs}), the \texttt{verify\_message} function performs a multi-stage validation:
\begin{enumerate}
    \item \textbf{Nonce Check:} Is the nonce fresh? (via \texttt{nonce\_manager.rs})
    \item \textbf{ZKP Check:} Does the ZKP prove ownership of the \texttt{did\_cid}?
    \item \textbf{Signature Check:} Does the signature match the content and the public key retrieved from the (now-trusted) \texttt{did\_cid}?
\end{enumerate}
Only if all three checks pass is the message accepted. This creates a trustless, authenticated broadcast channel for agents to perform tasks like service discovery.

\subsection{Iroh for High-Performance Direct Connection}
While Libp2p is excellent for discovery, it is not optimized for high-throughput, low-latency, 1-to-1 data exchange. For this, DIAP integrates Iroh, a next-generation P2P stack built on QUIC.

As implemented in \texttt{src/iroh\_communicator.rs}, Iroh is used for the second phase of communication: interaction after identity has been established.

The \texttt{iroh\_complete\_closed\_loop.rs} example demonstrates this perfectly:
\begin{enumerate}
    \item Two agents (Node 1, Node 2) establish their Iroh endpoints.
    \item Node 1 (receiver) calls \texttt{endpoint.accept()} to listen for connections.
    \item Node 2 (sender) uses the \texttt{NodeAddr} of Node 1 (which it would have learned via DIAP discovery) to call \texttt{endpoint.connect()}.
    \item A direct, QUIC-based connection is established.
    \item The agents open a bidirectional stream (\texttt{open\_bi()}) and exchange structured JSON-RPC messages.
\end{enumerate}
This dual-stack architecture gives DIAP a significant advantage: agents can use the robust, decentralized discovery of Libp2p to find each other, then switch to the high-performance, direct-connection model of Iroh to perform work.

\section{Demonstration \& Evaluation}
The DIAP SDK includes a comprehensive set of examples in the \texttt{examples/} directory that serve as both a demonstration of the protocol and an integration test suite.

\begin{itemize}
    \item \textbf{\texttt{complete\_auth\_demo.rs}:} This is the canonical example. It demonstrates the full end-to-end flow:
    \begin{enumerate}
        \item Two agents, "Alice" and "Bob", are created.
        \item Each agent generates a \texttt{KeyPair} and registers their identity to IPFS, receiving a \texttt{cid}.
        \item A \texttt{mutual\_authentication} is performed. Alice generates a ZKP proving ownership of \texttt{alice\_cid}. Bob verifies it.
        \item Bob generates a ZKP proving ownership of \texttt{bob\_cid}. Alice verifies it.
        \item Only when both verifications pass is \texttt{mutual\_trust} established.
    \end{enumerate}
    
    \item \textbf{\texttt{pubsub\_verification\_loop\_demo.rs}:} This example demonstrates the Libp2p broadcast stack. It shows agents registering their CIDs and then using the \texttt{PubsubAuthenticator} to send and verify ZKP-authenticated messages over a public gossip channel.
    
    \item \textbf{\texttt{iroh\_complete\_closed\_loop.rs}:} This example demonstrates the Iroh direct connection stack. It creates two Iroh endpoints, establishes a direct QUIC connection, and passes structured JSON request/response messages over a bidirectional stream, proving the viability of the high-performance stack.
    
    \item \textbf{\texttt{cross\_platform\_demo.rs}:} This example specifically tests the \texttt{UniversalNoirManager}. It confirms that the manager initializes on the default \texttt{Embedded} backend and can successfully generate and verify proofs, demonstrating the success of the zero-dependency migration.
\end{itemize}

\section{Conclusion}
We have presented the Decentralized Intelligent Agent Protocol (DIAP) and its accompanying Rust SDK. DIAP offers a novel solution to autonomous agent identity by binding identity to an immutable IPFS CID and using ZKP for stateless proof of ownership. This "ZKP-on-CID" approach overcomes the latency and reliability issues of traditional mutable identity systems like A2A.

Our implementation provides a robust, production-ready framework featuring a hybrid P2P communication stack (Libp2p and Iroh) for both discovery and interaction, and strong privacy-preserving features via encrypted P2P endpoint information.

Finally, we solved a major systems engineering challenge in ZKP deployment by creating a zero-dependency SDK. Through a universal backend manager and compile-time circuit pre-compilation, we have made advanced cryptographic identity accessible to developers without requiring external toolchains.

Future work will focus on expanding the capabilities of the Noir circuit, formalizing the security proofs of the protocol, and gathering user feedback from real-world DePIN and AI agent deployments to further optimize the P2P communication stack.



\end{document}